\documentclass{ijcaArticle}
\usepackage{url}
\usepackage{cite}
\begin{document}

\title{A Review of Network Evolution Towards a Smart Connected World} 

\author{ 
   \large Olivia Haring \\[-3pt]
   \normalsize School of Information Technology  \\[-3pt]
    \normalsize  University of Cincinnati\\[-3pt]
    \normalsize Cincinnati, OH, USA \\[-3pt]
  \and
   \large Sylvia Worlali Azumah\\[-3pt]
   \normalsize School of Information Technology  \\[-3pt]
   \normalsize  University of Cincinnati\\[-3pt]
   \normalsize Cincinnati, OH, USA \\[-3pt]
\and
   \large Nelly Elsayed\\[-3pt]
   \normalsize School of Information Technology\\[-3pt]
    \normalsize University of Cincinnati\\[-3pt]
    \normalsize Cincinnati, OH, USA \\[-3pt]
}

\terms{A Review of Network Evolution Towards a Smart and Connected World}

\maketitle

\begin{abstract}
With the rapid innovations in technology, wireless internet-connected devices are more ubiquitous than ever and can be found in virtually every aspect of both our personal and professional lives. In this paper, we propose a comprehensive literature review that focuses on various network components that create connectivity among different devices, specifically Wireless Sensor Networks (WSNs), Radio-Frequency Identification (RFID) tags, Internet of Things (IoT) devices, and how these devices helped usher in the 4th Industrial Revolution, or Industry 4.0. This paper focuses on the protocols, architecture, uses, security concerns, and solutions used in these network technologies, as well as their differences and similarities.
\end{abstract}

\section{Introduction}

The Internet of Things (IoT) is where the digital and physical worlds collide. IoT was first described as an internet-based information service. Much like how the internet revolutionized the way humans communicate with one another, IoT has helped usher in a new computing era. A report issued by Cisco estimates that there will be over 500 billion internet-connected devices that utilize sensors by 2030~\cite{RefWorks:doc:5fe9771d8f086f28cd458c62}.

From rudimentary wireless sensor networks (WSNs) to Radio-Frequency Identification (RFID), to the most complex of the Internet of Things (IoT) devices that form the backbone of smart factories, smart homes, and even entire smart cities, the technologies that underlie IoT devices have significantly changed the way humans interact with one another, their surroundings, and society as a whole. This comprehensive overview focuses on the history, application, architecture, challenges, solutions, and future of such technologies.  

In this paper, we provide a comprehensive exploration of the evolution of different devices' networks. We discuss the inception of wireless sensors and RFID technology, how these technologies are interconnected to form complex IoT systems, and what this means for future advancements in industrialization. 

\section{Wireless Sensor Networks}


Wireless sensor networks (WSNs) are comprised of many sensors that communicate by transmitting digital packets of information~\cite{tree2014wireless}. WSNs can contain anywhere from one sensor to hundreds of thousands of sensors. When these sensors are clustered together, they form what are referred to as “nodes.” These nodes are autonomous and extremely limited in their resources, due to the fact that they have minimal power supplies, processing capabilities, storage, etc. The lean nature of WSNs proves to be both an advantage and disadvantage~\cite{RefWorks:doc:5fac864ae4b014437ad91b13}.  

\subsection{History of WSNs}

The first utilization of wireless sensor networks began in the 1950s with the United States military and the utilization of the ``Sound Surveillance System" (SOSUS). This technology was used to detect enemy submarines and utilized acoustic sensors that measure the amplitude of ocean waves. Similar technology remains in use by the US military to this day~\cite{7868374,yick2008wireless}.  

\subsection{Architecture of WSNs}

Wireless sensor networks are comprised of many sensors that communicate by transmitting digital packets of information~\cite{raghavendra2006wireless}. WSNs can contain anywhere from one sensor, to hundreds of thousands of sensors \cite{yick2008wireless}. When these sensors are clustered together, they form what is referred to as ``nodes". These nodes are autonomous and extremely limited in their resources, due to the fact that they have minimalistic power supplies, processing capabilities, storage, etc. The lean nature of WSNs proves to be both an advantage and disadvantage \cite{perillo2005wireless}.  

\begin{figure}[htbp]
\centerline{\includegraphics[width=0.5\textwidth]{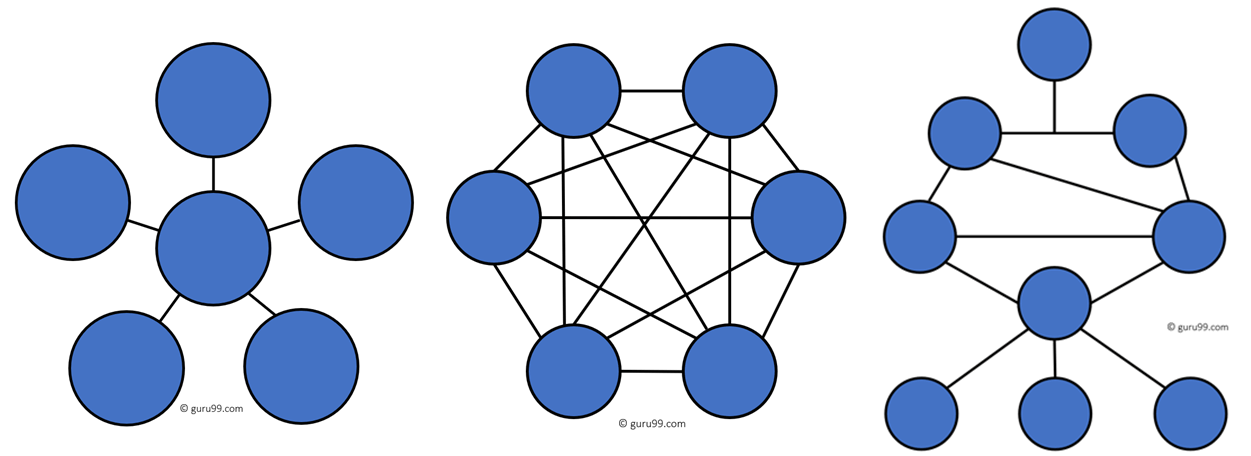}}
\caption{Example of a star, mesh, and hybrid topologies.}
\label{topology}
\end{figure}

WSN nodes are commonly arranged in a mesh, star, or hybrid (also know as tree) topology as shown in Figure~\ref{topology}~\cite{bruno2005mesh,baykas2014spectrum,singh2018survey, mamun2012qualitative}. These nodes collect information from individual sensors and transmit that information to other nodes~\cite{mcgrath2013key}. These sensor nodes are relatively simplistic when compared to today's modern technology and typically consist of a microcontroller, transceiver, external memory/storage, and at least one sensor~\cite{vieira2003survey}. However, it is common to have WSNs with hundreds, or even hundreds of thousands, of individual nodes.  The overall architecture of a WSN includes these sensor nodes, an end-user, and a backend infrastructure that allows the end user to access these sensor nodes~\cite{6805127}. The underlying architecture will vary depending on what the purpose of the WSN is~\cite{7868374}. 

\subsection{Types of WSNs}

While there are countless applications for individual sensors, there are five main types of WSNs: underground, underwater, terrestrial, multimedia, and mobile \cite{RefWorks:doc:5fac864ae4b014437ad91b13,singh2018survey}.

\subsubsection{Terrestrial WSNs}
In terrestrial WSNs, hundreds or thousands of wireless sensors are deployed in a specific area. The sensors are either distributed in an unstructured (ad hoc) or structured (preplanned) fashion. When distributed ad hoc, the sensors are randomly distributed within the target area. As the name implies, preplanned distribution involves advanced planning to ensure optimal placement of the sensor nodes.  

For terrestrial WSNs, the ability to reliably communicate and transmit data is crucial, especially in the oftentimes challenging environments they are deployed in. While battery power is limited in this type of WSN, it is possible to equip the sensor nodes with backup power supplies, such as solar cells \cite{RefWorks:doc:5fac864ae4b014437ad91b26,singh2018survey}.   

\subsubsection{Underground WSNs}
In underground WSNs, sensors are placed underground, most often in caves, mines, or simply burying them. Additional nodes are located above ground so that information can be transmitted from the sensors to the base station where the data is then collected and analyzed. Underground WSNs are more costly than terrestrial WSNs when it comes to their equipment, deployment, and ongoing maintenance. Since sensors in this type of deployment must reliably communicate through soil, rocks, water, and other earthen material, they incur increased costs. There is also the concern of signal loss because of the harsh underground environment. Unlike terrestrial WSNs, underground sensor deployment requires careful planning. As with terrestrial WSNs, underground sensor nodes have limited battery power and capacity, and once they are deployed underground, it is difficult and oftentimes impossible to repair, replace, or recharge sensor batteries~\cite{RefWorks:doc:5fac864ae4b014437ad91b26,singh2018survey}.  

\subsubsection{Underwater Wireless Sensor Network}
In underwater wireless sensor network (UWSN), sensor nodes and autonomous underwater vehicles are used to gather data from aquatic environments. This is also a very costly deployment type, and as a result, fewer sensor nodes are deployed compared to underground or terrestrial WSNs. In this deployment type, typical communication is accomplished via transmission of acoustic waves. Special challenges for underwater WSNs include limited bandwidth, transmission delays, signal feeding issues, and the high rate of sensor node failure as a result of harsh environmental conditions. Sensor nodes that are placed underwater must be able to configure themselves and adapt to the challenging oceanic environments in which they are deployed. Underwater sensor nodes also have very limited battery capacity which cannot be replaced or recharged once they are deployed, so battery conservation is critical in UWSNs~\cite{RefWorks:doc:5fac864ae4b014437ad91b26,singh2018survey}. 

\subsubsection{Wireless Multimedia Sensor Network}
The fourth type of wireless sensor networks are Wireless Multimedia Sensor Network (MWSN). This deployment type consists of a network of wireless, interconnected sensors that can retrieve multimedia content such as video, audio images, and sensor data from the environment in which they are deployed. This type of wireless sensor network consists of a number of low-cost sensor nodes equipped with cameras and microphones. Multimedia center nodes are deployed in a pre-planned fashion to ensure adequate coverage of the area being surveyed. There are a number of unique challenges to this type of WSN, including higher bandwidth requirements, higher energy consumption, and quality of service provisioning \cite{RefWorks:doc:5fac864ae4b014437ad91b26,singh2018survey}.  

\subsubsection{Mobile Wireless Sensor Network}
The final type of wireless sensor networks are mobile WSNs. This deployment type consists of a group of sensor nodes that are able to move independently and interact with their physical environment. These nodes have the ability to sense, compute, and communicate just as static nodes would. One difference is that mobile nodes are capable of re-positioning and reorganizing themselves within the network. A mobile WSN can be deployed in one way,then the  nodes can eventually spread out to gather information as needed. Information gathered within a mobile node can be communicated to another mobile node if they are within range of one another. Another key difference is that mobile WSNs use dynamic routing, as opposed to the fixed routing used within static WSNs \cite{RefWorks:doc:5fac864ae4b014437ad91b26,singh2018survey}.  

Common concerns and  challenges with mobile WSNs include deployment, navigation, coverage, maintenance, data processing, and battery life. 

Regardless of which type of deployment is used, wireless sensor networks all share several traits with one another: scalability, reliability, responsiveness, mobility, and power efficiency \cite{7868374}. All wireless sensor networks should have the capacity for scalability. Users should be able to expand the network and add or remove nodes as required, in a relatively easy fashion. Wireless sensor networks should also be generally reliable. There are many different methods for reducing the power usage of sensor nodes, which result in an increase in the lifetime of the network, and sensor consistency. Wireless sensor networks should also be responsive. Due to their simplistic architecture, wireless sensor networks should have a quick response time, even when things such as harsh environmental conditions are taken into consideration. Wireless sensor networks should also have a high degree of mobility, as this is the fundamental feature of WSNs. Since it is an inherently wireless network, mobility is an absolute necessity. Due to their deployment for long periods of time and the need for consistent and ongoing data transmission,power efficiency is crucial for wireless sensor networks ~\cite{kan2007accurate}. 

\subsection{WSN Communication Protocols}

There is not one single communication protocol that is universally used in the deployment and life cycle of wireless sensor networks \cite{7868374}. Instead, there are numerous protocols used at the transport, network/routing, datalink, and physical layers which are utilized depending on the purpose of the WSN in question. Due to the inherent restraints of WSNs, it is imperative to be mindful of energy consumption, latency, and load balancing when determining the appropriate protocol(s). In addition, emerging research has proposed cross-layer protocols to address various shortcomings in existing protocols. 

Historically, research has focused primarily on protocols concerning the network or routing layer because this is the layer that typically varies the most depending on the WSNs’ purpose. The three main types of network layer architectures are categorized as either data-centric, hierarchical, or location-based \cite{singh2010routing}. 

\subsubsection{Data-Centric}
In data-centric routing protocols, also known as flat-based routing, all nodes in the WSN have the same role in that they transfer data via flooding. A potential issue with this protocol is that implosion is possible, meaning that two nodes send similar packets, inadvertently consuming large amounts of energy, and thus shortening the life of the WSN~\cite{akkaya2005survey,singh2010routing}.

\subsubsection{Hierarchical-Based}
In hierarchical-based, also referred to as cluster-based, two network layers are utilized: one to select the head cluster, and the other to send the actual data. The primary goal of this protocol is to bundle the nodes in such a way that pre-processing of data can be performed, so as to reduce energy consumption~\cite{akkaya2005survey,singh2010routing}. 

\subsubsection{Location-Based}
In location-based routing protocols, the physical location where data originated from is used to transport information to a desired region or regions in the WSN, as opposed to sending it throughout the entire network. Except for the most simplistic, virtually all WSNs collect location information in order to calculate the distance between two nodes, and subsequently determine the energy usage~\cite{akkaya2005survey,singh2010routing}.  

\subsubsection{Data-centric}
In data-centric routing protocols, also known as flat-based routing, all nodes in the WSN have the same role in that they transfer data via flooding. A potential issue with this protocol is that implosion is possible, meaning that two nodes send similar packets, inadvertently consuming large amounts of energy, and thus shortening the life of the WSN~\cite{akkaya2005survey,singh2010routing}.  

\subsubsection{Physical Layer}
The physical layer is where tasks related to radio frequency and actual computations occur. Challenges in the physical layer include finding affordable radio transceivers that are energy-efficient, are not prone to significant radio interference, but also complex enough to perform the required tasks needed of the WSNs~\cite{fahmy2016protocol}.   

\subsubsection{Cross-Layer Protocols}
There has been a great deal of emerging research concerning cross-layer protocols for WSNs with the underlying goal of improving overall performance. These enhancements are often specific to the type of WSN and their precise priorities for improvement. Historically, there has been a focus on the interaction between the physical, data link, and routing layers.  For example, existing research on cross-layer protocols have been proposed for specific objectives such as maximizing successful packet transmission, sleep duration, throughput, minimizing energy consumption, or just optimizing the overall performance of a WSN~\cite{nedevschi2008reducing,resner2018design}.   

\subsection{Security Concerns and Solutions}
Due to the unique nature of wireless sensor networks and their overall simplicity, traditional cybersecurity measures and techniques are not necessarily possible or effective. WSNs are traditionally left unattended and often do not have predefined infrastructures. Once data is transmitted, it is very easy for a malicious actor to sniff or spy on network traffic.Oftentimes sensor nodes are not made tamper-proof due to strict budget requirements, sensor nodes are not always made tamper proof, and therefore have no protection against physical security attacks. One of the inherent security benefits of WSNs is that they can be deployed in very harsh environments that are not easily accessible to everyday people. Cryptographic algorithms are generally used to address common security issues such as confidentiality, integrity, authentication, and availability. However, these cryptographic solutions are oftentimes simplified due to the resource limitations of wireless sensor networks. This in turn reduces the effectiveness of these security measures. Because of this, it is recommended there be a second line of defense when deploying a wireless sensor network. These defenses can include elements such as intrusion detection systems, and trust and authentication models. However, one of the fundamental flaws in wireless sensor networks remains their limited battery capacity, which limits the level of sophistication that can be used in security solutions~\cite{RefWorks:doc:5fe0f4a48f0866e15553c7ab}.

\section{Radio Frequency Identification}
Radio frequency identification (RFID) is a type of wireless, automated technology that utilizes radio signals to identify tiny integrated circuit transponders known as RFID tags. These tags are equipped with antennas that communicate with their reading devices (RFID readers) using electromagnetic fields.  It is also common for there to be a back-end database that collects information related to the physical objects that have RFID tags on them.  

\begin{figure*}[htbp]
\centerline{\includegraphics[width= 11.5 cm, height= 6.5cm]{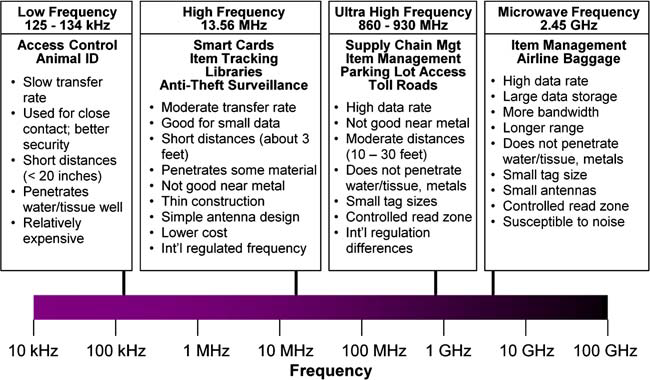}}
\caption{An overview of RFID frequencies and their real-world applications.}
\label{frequencies}
\end{figure*}

There are three commonly used types of RFID tags, each with their own advantages and disadvantages. The first is known as an active, or battery powered RFID tag. These tags need battery power to function,which makes them much more costly, and therefore less common. These tags typically have enough battery power to last several years, possess the ability to read and write data, and can transmit signals over the greatest distance~\cite{curtin2007making}.
The second type of RFID tag is known as a passive RFID tag. These tags do not contain a battery, but instead work by using electromagnetic energy that is transmitted from the reader to the tag. Since they do not contain a battery, they are much more cost efficient, economical, and often much smaller then active tags. 
The third type of RFID tag is referred to as a semi-passive tag. While these tags contain a battery that primarily functions to ensure data integrity, it is the signal sent from the reader that actually generates power which allows signal transmission from the tag to the reader~\cite{curtin2007making}.  
Passive RFID tags typically work on three different radio frequencies: low frequency, high frequency, and ultra-high frequency. High frequency communication is also known as Near Field Communication or NFC~\cite{weinstein2005rfid, ngai2008rfid, communications2005rfid}.   Figure~\ref{frequencies} shows the description of the RFID frequencies and their real-world applications.

\begin{figure}[htbp]
\centerline{\includegraphics[width=0.5\textwidth]{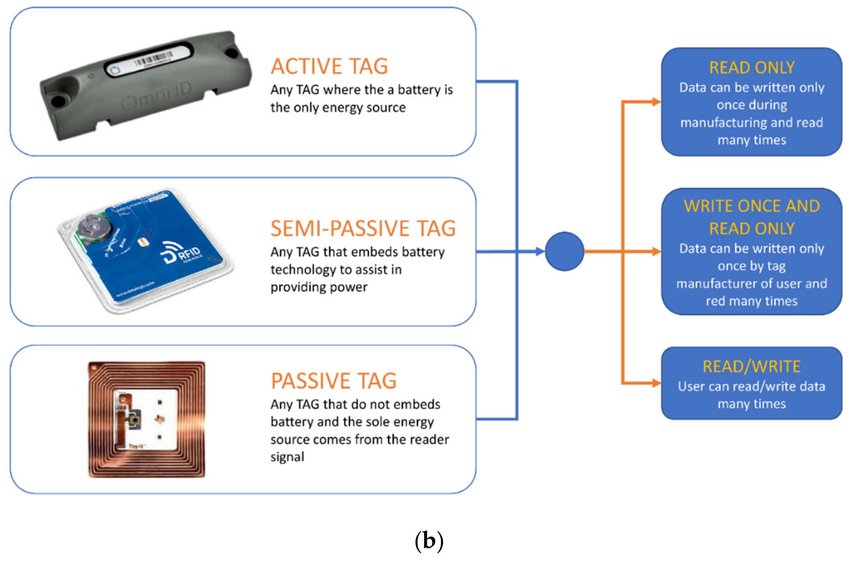}}
\caption{The three types of RFID technology.}
\label{RFID_technologies}
\end{figure}

\subsection{History of RFID}
A precursor to RFID technology was first observed in 1945 during WWII and was used by Soviet spies. This rudimentary technology rebroadcast radio waves with added audio information.  

The concept of RFID as we know it today was first envisioned by Harry Stockman in his 1948 paper ``Communication by Means of Reflected Power." However, it would take more than 30 years for Stockman’s vision to become technologically feasible.  

By the 1970s, corporations had made considerable advancement in the development of RFID technology. Notable examples include Raytheon's Raytag in 1973 and Richard Klensch of RCA developing an electronic identification system in 1975. 

The 1980s is when RFID technology began to actually be implemented. However, the means by which RFID was implemented varied. In the United States, there was a focus on implementing this technology for transportation, employee and personnel access, and to a lesser degree, animal monitoring. In Europe, there was a larger focus on RFID for animal monitoring. Despite the lesser focus on transportation applications, the European country of Norway became the first to implement RFID technology as a means of collecting tolls. This was followed closely in the United States in 1989 when the Dallas North Turnpike became the first to use RFID for toll collection. Soon after, the Port Authority of New York and New Jersey implemented RFID-based toll collections for buses going through the Lincoln Tunnel.  

The 1990s saw an increased implementation of RFID as a means of toll collection throughout the United States, and there were over three million RFID tags installed on rail cars in North America by the end of the decade.  

The $\mathrm{21^{st}}$ century ushered in a new era for RFID technology, as major advances resulted in the smallest ever RFID technology-- so small it could be manufactured as a thin, adhesive label that could be attached to virtually any surface. Connecting this breakthrough technology to the internet is what ultimately helped RFID become a critical component of numerous IoT devices~\cite{RefWorks:doc:5fe0fb878f080f94e0b5b137,landt2005history}.

\subsection{Uses for RFID Technology}
Due to passive RFID tags costing only cent to manufacture and having the ability to be applied virtually anywhere in the form of stickers, this technology is a cost-effective way to turn just about any item into a ``smart" item. Much like wireless sensor networks, RFID tags are particularly useful in challenging environments since they require little to no human supervision once they are deployed. RFID tags can typically be read through a number of environmentally challenging conditions, such as snow, ice, fog, or when objects cannot be directly touched, such as items on pallets in warehouses. Oftentimes, readers do not need to have direct contact with the RFID tag in order to be read. This is beneficial in occupational situations where the object in question may be exposed to paint, grime, mud, etc. 

RFID technology is seen in virtually every industry imaginable, from inventory control systems in retail stores, toll collection, logistics and supply chain management, animal tracking, employee identification badges, healthcare, consumer smart devices, and more. RFID tags can even be implanted into the human body and were approved by the FDA in 2004~\cite{RefWorks:doc:5fac7e04e4b014437ad919a8,dobkin2012rf,ngai2008rfid}.  

\subsection{RFID Communication Protocols}

RFID tag technology utilize low frequency (125–134 kHz), high frequency (13.56 MHz), and ultra-high frequency (300 MHz-3GHz) radio signals to communicate wirelessly with a reader. Figure~\ref{RFID_technologies} shows the three types of RFID radio signals to communicate wirelessly with a reader technology types. Unlike WSNs, the various types of RFIDs have well-established industry standards regarding communication protocols~\cite{ibrahim2019review,communications2005rfid}.

The most commonly utilized standard for high frequency RFID tags is ISO/IEC 15693, which was most recently updated in 2019. This standard is applicable to what are commonly referred to as “vicinity cards” or RFID tags that have a maximum read distance of 1 meter. As with the standard for LF RFID tags, ISO/IEC 15693 provides precise technical parameters for HF RFID tags, including the physical layer used between the reader and tag, and the anti-collision methodology used to detect and communicate with a specific tag when several tags are present.  

The most commonly utilized standard for high frequency RFID tags is ISO/IEC 15693, which was most recently updated in 2019. This standard is applicable to what are commonly referred to as “vicinity cards” or RFID tags that have a maximum read distance of one meter. As with the standard for LF RFID tags, ISO/IEC 15693 provides precise technical parameters for HF RFID tags, including the physical layer used between the reader and tag, as well as the anti-collision methodology used to detect and communicate with a specific tag when several tags are present.  

While both active and passive tags operate on the ultra high frequency range (300 to 100MHz), only two frequency ranges are actually utilized. These ranges include 433 MHz for active tags, and a range of 860-890 for passive tags. Unlike WSNs, passive UHF RFID tags have a widely accepted industry standard known as the Electronic Product Code (EPC) Class1 (C1) Generation2 (Gen2) standard, informally referenced as EPC Gen 2~\cite{al2008challenges,adhiarna2009standardization, RefWorks:doc:5fac8713e4b08d77ad27eab4, dobkin2012rf}.

\subsection{RFID Security Concerns and Solutions}
As with any ubiquitous technology, there are numerous security concerns. These include concerns regarding consumer privacy, physical tampering, and malicious cyberattacks.  

\subsubsection{Consumer Privacy}
Since RFID technology is essentially undetectable, this can result in both profiling and location tracking of consumers without their consent or knowledge. Existing publications and literature regarding consumer privacy concerns of RFID technology can be categorized into five general themes. The first is the undetectable and concealed nature of RFID tags. Tags can be embedded in or on almost any object without the consumer knowing. Another criticism is that RFID technology provides the ability to mass-identify objects. Each RFID tag consist of unique identifying information. In a worst-case scenario, this could lead to the creation of a globalized system in which every physical object is identified and linked to its owner at the point of sale, or when it is transferred. There is also concern regarding the inherent ability to collect massive amounts of data. Since a primary function of RFID data that is collected is the creation of databases, databases containing this tag data could be linked to personally identifying information. This particular concern is especially worrisome as computing power continues to increase. RFID tags also allow the opportunity to profile and track people. If personally identifying information were to be linked with unique information contained in RFID tags, individuals could be tracked or profiled without their consent or knowledge. Finally, the ability for tag readers to function without being directly in contact with tags has created additional privacy concerns. Readers could be incorporated into virtually any setting where people gather, resulting in information being easily accessible and collected into databases~\cite{RefWorks:doc:5fac7e04e4b014437ad919a8}.

\subsubsection{Physical Tampering}
Passive RFID tags inherently have poor physical security and are extremely prone to physical manipulation. This can include things such as attacks that permanently or temporarily disable the tags, or physically removing or destroying the tag. When RFID tags are used as anti theft measures in retail settings, it is possible for malicious customers to simply remove the RFID tag and walk out of the store with the merchandise.  

While RFID tags can be used in challenging environments, they are still susceptible two possible destruction as a result of extreme environmental conditions, such as temperatures that are too high or too low, or damage caused by rough handling. Active RFID tags can also be made inoperable by merely removing or discharging the battery. RFID tags are very sensitive to static electricity, and their electronic circuits can be instantly damaged by electrostatic discharge . This is especially concerning in warehouse environments, as conveyor belt picking systems often carry a large amount of static electricity. There are also several special privacy protecting devices that people can purchase, or create themselves, such as ``RFID zappers." 

There are also security concerns surrounding RFID readers. Handheld RFID readers can be destroyed, removed, or even stolen if they are left unattended. RFID readers oftentimes include sensitive information such as keys and other cryptographic credentials. Theft of RFID readers could potentially allow malicious attackers to gain access to the back-end database where sensitive information such as personally identifiable information or company trade secrets may be stored~\cite{RefWorks:doc:5fe0fd458f082ee73fb7be47}.

In order to safeguard RFID systems against low-tech attacks such as permanently or temporarily disabling tags, traditional countermeasures should be used, such as increased physical security with guards, fences, gates, locked doors and cameras. Whenever possible, it is advisable to not merely stick an RFID tag directly on an item, but embedded in the item and or packaging itself to prevent tampering. In retail settings, stores will often have alarm systems that are triggered if a tag is not deactivated at the point of sale~\cite{RefWorks:doc:5fac7e04e4b014437ad919a8}.  

\subsubsection{Cyberattacks}
Tags with little to no protection are especially vulnerable to eavesdropping, traffic analysis, spoofing, denial of service, and other cyber attacks. It is possible to initiate a denial of service attack by flooding an area with radiofrequency energy, thereby incapacitating RFID readers. RFID tags are inherently designed to be readable by any compliant reader. This in theory could allow any user with a reader to scan tagged items, often from significantly far away, potentially releasing sensitive information to malicious actors.  

It is also possible to mimic genuine RFID tags by writing correctly formatted data onto blank RFID tags. This could allow for spoofing of data. Malicious actors could also flood a system with an overwhelming amount of data, more than it was designed to handle. Theoretically, a person could remove a tag, and then place it on other items, causing the system to record useless data, and thereby devaluing the back end database. This is especially concerning for larger corporations that widely use RFID tags for inventory control and security measures, such as retail stores.  

Unauthorized readers can impact privacy by accessing tags that are lacking access control. Even if the tag content is secure it can still be tracked by the predictable tag responses; ``location privacy" can be affected by a traffic analysis attack. Attacker can also threaten the security of systems, which depends on RFID technology, through the denial of service attack~\cite{RefWorks:doc:5fac8713e4b08d77ad27eab4}.  

\begin{figure}[htbp]
\centerline{\includegraphics[width=8.5 cm, height=5cm]{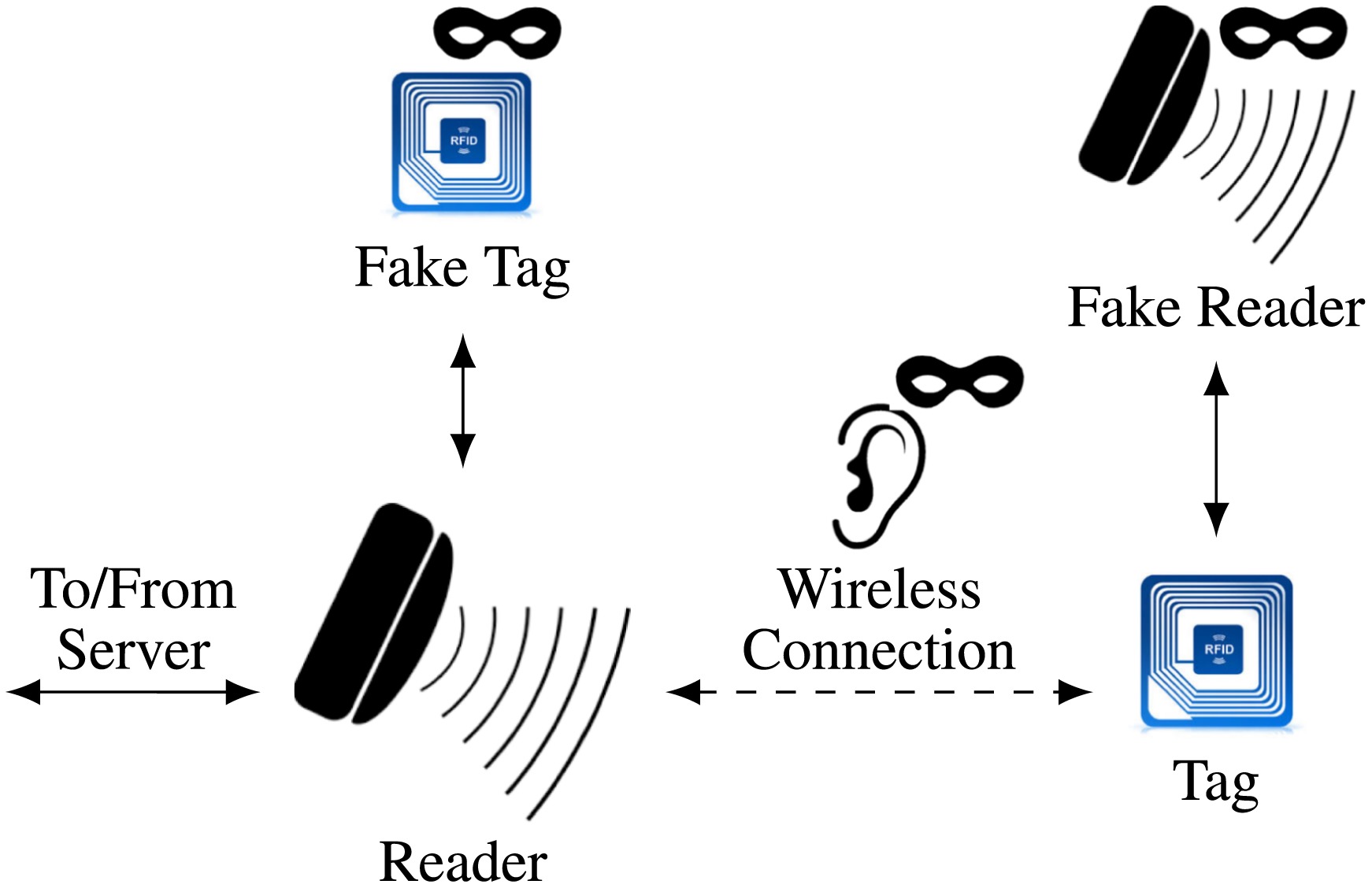}}
\caption{Fake RFID tags and readers are just one of the many security concerns surrounding this technology.}
\label{fake}
\end{figure}

RFID tag standards incorporate a 64-bit region that cannot be modified and remains unique to the tag itself. This can be used to authenticate the tag and defends against tag spoofing. A number of different attacks,such as replay attacks, can be rebuffed through the use of hidden authentication schemes such as serial numbers, or rudimentary key cryptography specific to RFID tags that has been developed by researchers~\cite{pateriya2011evolution, RefWorks:doc:5fac8713e4b08d77ad27eab4}.

RFID tags are indiscriminate, they are designed to be readable by any compliant reader. Unfortunately, this lets unauthorized readers scan tagged items unbeknownst to the bearer, often from great distances. 

Attackers and anti-RFID activists can mimic authentic RFID tags by writing appropriately formatted data on blank RFID tags, and could also remove RFID tags and plant them on other items, causing RFID systems to record useless data, discrediting and devaluing RFID technology~\cite{RefWorks:doc:5fe0fb878f080f94e0b5b137}. 

\section{Internet of Things}
The Internet of things or IoT is a system of internet-connected objects (or things), that are embedded with sensors or other data-collecting technology that enables them to send and receive data~\cite{al2017internet, RefWorks:doc:5fe9774e8f0896665637d1de}.  

\subsection{History of IoT}
The term ``Internet of Things" is credited as being coined in 1999 by Kevin Ashton, an employee of the Massachusetts Institute of Technology. He described it as ``a system of interconnection between the physical world and the Internet through the use of RFID and pervasive sensor devices that identify and observe the real world."  

However, a rudimentary interconnection between everyday objects and the internet had already been developed in the early 1980s. Employees at Carnegie Mellon University connected a soda vending machine to the Internet in order to check the inventory and availability of drinks in the machine~\cite{suresh2014state}.     

\subsection{IoT Architecture and Communication Protocols}
There is no single or general agreement about the architecture of IoT that is recognized by world's researchers and professionals. Various IoT architectures have been proposed which range from three layers to more complex architectures with seven or more layers. In each of these layers, you will find previously discussed protocols such as those found within WSNs and/or RFID, since IoT devices often contain these technologies.   

\begin{figure}[htbp]
\centerline{\includegraphics[width=8.5 cm, height=6cm]{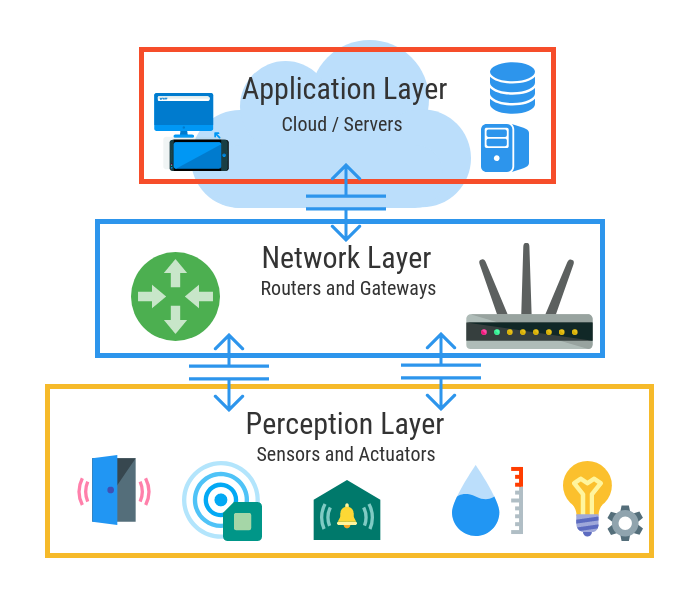}}
\caption{The three layer model of IoT architecture.}
\label{three_layered IoT}
\end{figure}

At its most simplistic form, there are three layers to IoT architecture: the perception, network, and application layers as shown in Figure~\ref{three_layered IoT}. 

\begin{figure}[htbp]
\centerline{\includegraphics[width=8.5 cm, height=5cm]{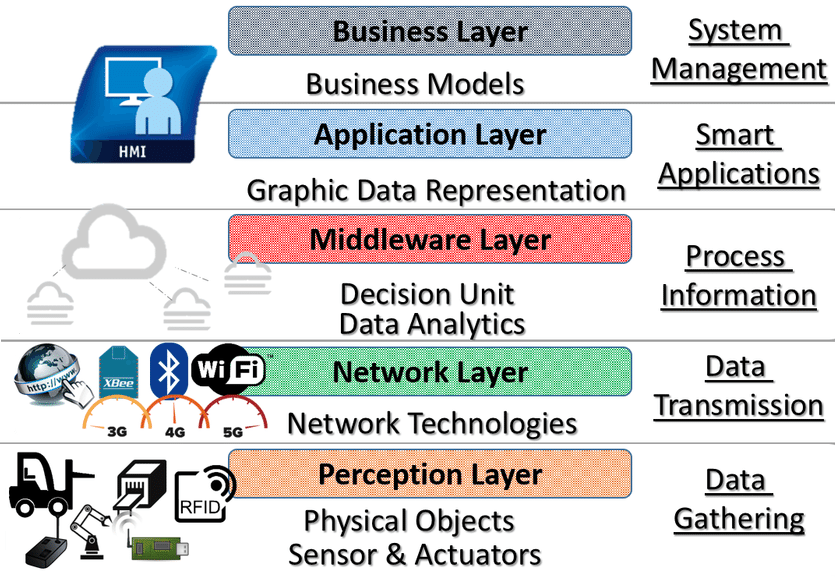}}
\caption{The five layer model of IoT architecture.}
\label{five_layersIoT}
\end{figure}

The five-layer IoT architecture consists of the following layers: perception, transmission/network, middleware, application, and business \cite{RefWorks:doc:5fe976898f083a89f269c228,kumar2018internet}. The five-layer IoT architecture is shown at Figure~\ref{five_layersIoT}.

\subsection{IoT Security Concerns and Solutions}
The fact that there is an overall lack of standardization and regulation around IoT security is itself a major security concern. Many IoT devices are composed of rudimentary technology such as RFID and WSNs, meaning that they lack the hardware capacity to support a vigorous security infrastructure.  

Especially in the context of industrial IoT such as factories, manufacturing plants, or general corporate settings, smart devices can prove to be extremely lucrative targets for threat actors. These can be targets for professional cyber criminals with significant training, which can result in severe financial damages and major consequences for the victim organizations. 

For all IoT devices, but especially those that are particularly sophisticated or contain sensitive or valuable data, it is advisable to implement a number of cybersecurity best practices. These can include things such as two factor authentication, biometrics, digital certificates, and ensuring there are antivirus, antimalware, firewalls, and intrusion prevention and detection systems present. It is also advisable that all data between IoT devices and back-end systems be encrypted ~\cite{RefWorks:doc:5fe0fd458f082ee73fb7be47, RefWorks:doc:5fe976898f083a89f269c228, RefWorks:doc:5fe0fe8b8f082ee73fb7be7d, RefWorks:doc:5fe0fe8b8f082ee73fb7be7c}.    

\section{Industry 4.0} 
 The Fourth Industrial Revolution, also known as Industry 4.0, is a term coined to refer to the rapid transformation into automation of traditional manufacturing and industrial practices the smart technology. Industry 4.0 has led to the creation of  what is known as "smart factories." Technologies such as machine-to-machine communication are integrated with the Internet of Things and manufacturing processes to increase automation, improve communication, and increase the self-monitoring, analyzing, and diagnosing of machinery without the need for human intervention. Technology associated with Industry 4.0 is heavily reliant upon cyber-physical systems such as sensors that collect and analyze huge amounts of data which is then used by machine operators, manufacturers, and other stakeholders to improve efficiency and output. Advancements in computing power and processing speed allows for systems which can scan huge sets of data and produce insights that can be acted upon quickly by humans. Industry 4.0 and Big Data go hand-in-hand, as these technologies allow collection of data at scales never before thought possible~\cite{lasi2014industry, RefWorks:doc:5f89f5ffe4b093c4e9de4932, RefWorks:doc:5fe9774e8f0896665637d1de}.

\section{Conclusion}
While there have been vast technological breakthroughs regarding IoT technology, we have only scratched the surface of what this technology is truly capable of. IoT is like the Wild West in that it is largely unregulated, with every company hoping to strike gold. There remain a number of concerns that need to be resolved for this technology to continue advancing. These issues include security and privacy, storage, energy usage, and the communication, compatibility, and standardization between different IoT devices.

\section*{Acknowledgment}
This material is based upon work supported by the National Science Foundation under Grant No. (CNS-1801593). Any opinions, findings, and conclusions or recommendations expressed in this material are those of the author(s) and do not necessarily reflect the views of the National Science Foundation.

\bibliography{references}
\bibliographystyle{ieeetr}

\end{document}